\documentclass[11pt]{revtex4-1}
\usepackage{geometry}
\usepackage{graphicx}
\usepackage{amsmath}
\usepackage{amssymb}

\newcommand{\be}{\begin{equation}}
\newcommand{\ee}{\end{equation}}
\newcommand{\ba}{\begin{eqnarray}}
\newcommand{\ea}{\end{eqnarray}}
\newcommand{\la}{\lambda}

\newcommand{\Tr}{\rm Tr}

\newcommand{\si}{\sigma}
\newcommand{\vf}{\varphi}
\newcommand{\G}{\Gamma}

\begin{document}

\title{The crossover region between long-range and short-range interactions for the critical exponents }

\author{E.~Brezin$^1$, G.~Parisi$^2$ and F.~Ricci-Tersenghi$^2$}

\affiliation{
$^1$ Laboratoire de physique th\'eorique, \'Ecole Normale Sup\'erieure, 24 rue Lhomond, 75005 Paris, France\\
$^2$ Dipartimento di Fisica, INFN -- Sezione di Roma 1, CNR -- IPCF UOS Roma, Universit\`{a} ``La Sapienza'', P.le A. Moro 5, I-00185 Roma, Italy
}

\begin{abstract}
It is well know that systems with an interaction decaying as a power of the distance may have critical exponents that are different from those of short-range systems. The boundary between long-range and short-range is known, however the behavior in the crossover region is not well understood. In this paper we propose a general form for the crossover function and we compute it in a particular limit. We compare our predictions with the results of numerical simulations for two-dimensional long-range percolation.
\end{abstract}

\maketitle

\section{Introduction}

It is well known that critical exponents have a  high degree of universality. Quite different systems belong to the same  large universality class. The critical exponents are usually determined by the symmetries of the problem and by the dimension of the space. 

These results are valid for short-range interactions. When the interaction decays slowly at large distance the situation changes: in the case of an interaction decaying as a power law,  the critical exponents may depend on the exponent characterizing the power decay of the interaction. This is a  well studied phenomenon. One finds that, if the  interactions decay as a power of the inverse of the distance, provided the power is sufficiently large, the critical exponents are the same as those of the short-range model, while for slowly decaying interaction the critical exponents do depend on that  power law.

Therefore there is a  value of the exponent that separates the short-range behavior from the long-range behavior. In this paper we are interested in understanding what happens at this crossover point. We will find that exactly at this point there are logarithmic corrections to the the standard critical exponents. We are interested also to study the whole  crossover region.

For definiteness let us consider a ferromagnetic  system with  a long-range  interaction in dimension $d$. Although our arguments are quite generic we restrict ourselves to a system with an internal $O(N)$-symmetry with a two-spin  interaction that decays as $1/r^{d+\sigma}$, with positive $\sigma$.  The value of $\sigma$ must be positive in order to have a well-defined thermodynamic limit. 

The behavior  of the critical  exponents $\nu$ and $\eta$ as function of $d$ and $\sigma$ has been studied in great detail. It is clear that for $\sigma$ large enough, the critical exponents  become equal to the short-range ones (i.e. $\nu_{SR}$ and $\eta_{SR}$). In this respect the most important results  have been obtained  by Sak \cite{Sak1,Sak2} using Wilson original approach. The main results are the following \cite{Sak1,Sak2}:
\begin{itemize}
\item For $\sigma>\sigma^*\equiv2-\eta_{SR}$,  the critical exponents have the value $\nu_{SR}$ and $\eta_{SR}$ as in short-range models.
\item For $ \sigma<d/2$ mean field theory is correct and  the critical exponents  have the mean field value $\nu=1/2$ and $\eta=0$.
\item The non trivial region is for $d/2<\sigma<2-\eta_{SR}$: this interval is non empty only if $d<4$. Here the critical exponent  $\eta$ is given by $2-\sigma$, while the exponent $\nu$ is a non-trivial function of $\sigma$ and $d$. The exponent $\nu$ can be computed in perturbation theory in various limits \cite{AAA}.
\end{itemize}

In other words in dimensions $d<4$ the value of $\sigma^*\equiv2-\eta_{SR}$ separates the region where $\eta$ has a simple value (i.e. $2-\sigma$), from the region where $\eta=\eta_{SR}$. 

The aim of this paper is to study in detail what happens in the region of $\sigma$ near  $\sigma^*$.  In particular we are interested in studying the spin-spin correlation function { at the critical point} for large space separation $x$ and small $\delta\equiv\sigma-\sigma^*$. 
We will argue  that in this region the following formula holds:

\begin{equation}
C(x,\delta)=\delta \, x^{-d+\sigma^*+\delta}  F(\delta \log(x))\equiv \delta \, x^{-d+\sigma}  F(\delta \log(x))\, ,
\end{equation}
where
\begin{eqnarray}
F(z) \approx -F^- \mbox{\ \ for \ \ } z\to -\infty \, ,  \label{I0} \\ \nonumber
F(z) \approx F^0/z \mbox{\ \ for \ \ } z \to 0 \, ,
\\ \nonumber 
F(z) \approx F^+ \exp(-z)  \mbox{\ \ for \ \ } z\to +\infty \, .
\end{eqnarray}
In other words in the large $x$ region we have:
\begin{eqnarray}
C(x,\delta) \approx - \delta\ F^-  x^{-d+\sigma} \mbox{\ \ for \ \ } \delta<0  \, , \label{I1}
\\ \nonumber 
C(x,0) \approx \frac{F^0 x^{-d+2-\eta_{SR}}}{\log(x)} \mbox{\ \ for \ \ } \delta= 0 \, ,
\\ \nonumber \,
C(x,\delta) \approx   \delta\ F^+  x^{-d+2-\eta_{SR}}  \mbox{\ \ for \ \ } \delta>0 \, .
\end{eqnarray}
Similar formulae can be derived in the small momentum region at the critical point.

A simple example of such a behavior is given by
\be
C(x,\delta) =x^{-d+2-\eta_{SR}}\frac{\delta }{x^\delta-1} \,,
\ee
where the crossover function is $F(z)=(\exp(z)-1)^{-1}$.

These formulae are interesting for two reasons:
\begin{itemize}
\item The surprising presence of a logarithmic correction with an universal exponent in the spin-spin correlation function.
\item Recently, attempts have been done to verify numerically the correctness of the theoretical prediction; without taking into account the crossover function and the results were not clear-cut \cite{I2,Picco,APR}. If the system size is not very large, a logarithmic correction may be mistaken for a power correction with a small exponent.
\end{itemize}

In this article we will present some arguments that suggest the correctness of the previous equations. We will not present a complete derivation of these results, but we believe that the arguments are compelling.

\section{The scaling relations}
\subsection{General considerations}
For convenience we will rewrite  in momentum space the  equations (\ref{I0}) and (\ref{I1}) of the introduction. We know that for small momentum $k$  exactly at the critical point we have
\begin{eqnarray}
G(k,\delta) \propto 1/k^\sigma \mbox{\ \ for \ \ } \delta<0  \, ,
\\ \nonumber 
G(k,\delta) \propto 1/k^{2-\eta_{SR}}  \mbox{\ \ for \ \ } \delta>0 \, .
\end{eqnarray}

It is natural to conjecture that in the crossover region at small $k$ the following scaling relation is satisfied: 
\begin{equation}
G(k,\delta)\approx \delta^m \mathcal{H}(\log(k) \delta)\, , \label{Crossing}
\end{equation}
where the exponent  $m$ is not a priori fixed: the computation of $m$ is one of the main results of this paper. 

The previous formulae imply that in the region of small $k$  we  have two different powers in $k$ for positive and negative $\delta$ . This kind of scaling relation is quite common in  crossover regions and we take it for granted. Its failure would be rather surprising.

In order to determine the behavior in the crossover region, we investigate  $G(k,\delta)$ for small $\delta$ in the limit where $k$ goes to zero. If we were able to find that for   negative $\delta$ (i.e. in the long-range regime where $\eta=2-\sigma$) we have
\begin{equation}
G(k,\delta)\approx |\delta|^s /k^\sigma \equiv |\sigma-\sigma^*|^s /k^\sigma\, , 
\end{equation}
we could safely conclude that $m=s$. In this way we have completed our task of determining the scaling form.

In the rest of the section we will present arguments pointing toward $m=s=1$. 
Let us consider what happens in perturbation theory near the critical point, in a theory with a dimensionless coupling constant $g$. At  zeroth-order in perturbation theory the propagator is given by
\begin{equation}
G(k)=\frac{1}{k^{\sigma}} \, .
\end{equation}
When loop corrections are added one finds
\begin{equation}
G(k)=\frac{1}{Z(g(k),\sigma) k^{\sigma}} \, ,
\end{equation}
$g(k)$ being the running coupling constant. It is well known that $Z(g(k),\sigma)$ is not divergent in perturbation theory for $0<\sigma<2$ ; indeed in this range the critical self-energy behaves for small $k$ as $k^2$ and the mean-field relation $\eta = 2-\si$ is not renormalized \cite{AAA}. Let us call $Z(\sigma)$ the value of $Z(g(k),\sigma)$ at the infrared fixed point.

Let us use the notation $\sigma=d/2-\epsilon$. We can extend the standard $(4-\epsilon)$-expansion for short-range models  to this case.
Indeed in perturbation theory we can construct the fully renormalized theory at $\epsilon=0$. By considering a simultaneous expansion in $d$ and in the coupling constant, we can derive the $\epsilon$ expansion for the exponents.

In the renormalized theory  the correlation functions are defined as
\begin{equation}
G_{R}(k)=G(k)/Z\ \ \ \\  \ \left.\left( k^{\sigma} G_{R}(k)\right)\right|_{k=M},
\end{equation}
$M$ being a given renormalization point.

The renormalization constant $Z$ is finite, so that we do not need do introduce it in an explicit way: in any case no anomalies are involved and no $\gamma$ function is needed  in the renormalization group approach. The finiteness of $Z$ implies that $\eta$ is not renormalized in the $\epsilon$-expansion.

We will argue that $Z$ should go to infinity when  $\sigma=\sigma^{*}(d)$. Indeed the renormalization constant $Z$ is infinite for short-range interactions and in the renormalization group equation one has to introduce a non trivial $\gamma(g)$-function, $g$ being the effective coupling constant. Therefore the two regimes, short and long-range exponents, are characterized respectively by a divergent and a finite $Z$.  We will show that $Z^{-1}$ must vanishes linearly at $\sigma^{*}(d)$.
We will present a general argument in the framework of the conformal bootstrap equations. We will explicitly verify the correctness of this assumption in the framework of the $1/N$ expansion.

\subsection{The  bootstrap equations}

Let us consider a  short-range Landau-Ginzburg model with a $\phi^4$-interaction. In general the two-point and the four-point correlation functions satisfy a complex set of non-linear integral equations. 

At the critical point we can use scale invariance to find the solution of these equations.  The coefficients of the terms present in the Landau-Ginzburg  Hamiltonian appear  in the equations and  they scale in a different way from the correlations functions at the critical point. Therefore the scale invariant part of the correlation functions {\sl at the critical point} satisfies equations that do not contain the parameters of the  effective Hamiltonian. This fact is at the basis of the universality of the critical exponents in short-range models.

On the contrary, in the case of long-range models,  in the scaling regime, the equation for the two-point correlation function contains the terms present in the the Hamiltonian, because all  terms scale in the same way:
\begin{equation}
G^{-1}(k,\sigma)=A k^\sigma +\Sigma(k,\sigma) \,,
\end{equation}
where $A$ is the term appearing in the Hamiltonian and $\Sigma$ is the self energy. All the terms in the previous equation scale as $k^\sigma $.

If we introduce the renormalized field, such that the renormalized correlation function $G_R(k)$ is given by
\begin{equation}
G_R^{-1}(k,\sigma)\equiv Z G^{-1}(k,\sigma)= k^\sigma\, ,
\end{equation}
we find that 
\begin{equation}
G_R^{-1}(k,\sigma)=Z A k^\sigma +\Sigma_R(k,\sigma)\, .
\end{equation}
Scale invariance implies that  in the region of small $k$ $\Sigma_R(k,\sigma)=\Sigma_R(\sigma)k^\sigma$, so that the previous equation reduces to
\begin{equation}
1=A Z+\Sigma_R(\sigma)\, .
\end{equation}
where $\Sigma(k,\sigma)=Z^{-1} k^\sigma \Sigma_R(\sigma)$. 

The  equations for the short-range models are very similar: they are obtained from the long-range ones simply by suppressing an explicit term proportional to $k^{\sigma}$ in the equation for the inverse propagator. 

The equations for  higher order correlation functions are the same for short and long-range models. Now the crucial quantity $\Sigma_R(\sigma)$ has no reason whatsoever to have any particular property at $\sigma=\sigma^*=2-\eta$. apart from 
\begin{equation}
\Sigma_R(\sigma^*)=1 \, ,
\end{equation}
 that is the condition that it should satisfy in the short-range model. In the generic case we expect that 
\begin{equation}
\left. \frac{d\Sigma_R(\sigma)}{d \sigma}\right|_{\sigma=\sigma^*}\ne 0
\end{equation}
and therefore we expect the $Z^{-1}$ vanishes linearly at $\sigma=\sigma^*$.

In the next subsection we  verify that these expectations are correct in the large $N$ limit,  at first order of the $1/N$ expansion. This is important since it gives  confidence in the soundness of the whole approach.

\subsection{The $1/N$ expansion}

\emph{The long-range spherical model}
\vskip2mm

Le us write the Landau-Ginzburg Hamiltonian for an $N$-component field $\phi_a(x)$, with $a=1,N$. We start with the $O(N)$-symmetric weight ($e^{-S}$) , where $S$ is given by:
\be S(\vec{\phi}) = \int d^d x \left[ \frac{1}{2} {(\nabla^{\si/2}\vec\phi)} ^2+ \frac{1}{2} r (\vec\phi)^2 + \frac{g}{8N} ((\vec\phi)^2 )^2\right] \ee
and  the temperature enters linearly only in the value of the parameter $r$. 

We will follow the well-known technique of the large-$N$ limit, as in Zinn-Justin \cite{ZJ}.
We first introduce the auxiliary imaginary field $\la(x)$ conjugate to $(\vec{\phi})^2$ :
\be 
\int D\la \exp\left({\int d^dx \left( \frac{N}{2g} \la^2 - \frac{Nr}{g}  -\frac{1}{2} \la \vec\phi^2\right)}\right)\propto  \exp \left({-\int d^dx \left[  \frac{r}{2} \vec\phi^2 + \frac{g}{8N}(\vec\phi^2)^2 \right]}\right) \, .
\ee
We then integrate on $(N-1)$ tranverse components of $\vec\phi$, along some direction (e.g. fixed by a vanishing external field) and we keep the longitudinal component that for definiteness we assume to be the first component $\phi_1$.  We finally set $\phi_1=\sqrt{N} \varphi$. We arrive to a reduced Hamiltonian:
\be S(\vf,\la) = N \int d^dx \left[ \frac{1}{2} (\nabla^{\si/2}\vf)^2 +\frac{1}{2} \la\vf^2+ \frac{r}{g} \la- \frac{1}{2g} \la^2\right] + \frac{(N-1)}{2} \Tr\log[- \Delta^{\si} + \la]\, .     \ee

The large $N$ limit is thus given by the saddle point equations in the two fields $\vf$ and $\la$ and the corrections are the usual loop expansion. In the absence of space varying external field,  we obtain the equations
\ba \vf\la &= & 0\nonumber\\  
\la -r -\frac{g}{2}\vf^2&=& \frac{g}{2} \int \frac{d^d q}{(2\pi)^d}\frac{1}{ q^\si +\la}\, , \ea 
 where   we suppose  that an ultraviolet  cut-off (e.g. the lattice) is present.

The previous equations can be solved easily. We find:
\begin{itemize}
\item Below $T_c$ (i.e. for $r<r_c $) the spontaneous magnetization $\vf$ does not vanish, thus we have 
\ba \la &=& 0\, ,\nonumber \\
 \vf^2 &=&- \frac{2r}{g} - \frac{1}{(2\pi)^d}\int \frac{d^dq}{q^{\si} }\, .\ea
 The massless Goldstone modes are here well-defined  only for $d>\si$. 
The saddle-point requires  $r<r_c$ with
 \be- \frac{2r_c }{g} = \frac{1}{(2\pi)^d} \int \frac{d^dq}{q^{\si} }\,, \ee
 i.e. in term of the reduced temperature $t$ proportional to $(T-T_c)$
 \be r-r_c = \frac{g}{2} t \,,\ee
 \be \vf^2  = -t \,.\ee
 The exponent $ \beta$ (for the spontaneous magnetisation) remains equal to the usual long-range value $1/2$. 
 \item Above $T_c$ the magnetization $\vf$ vanishes and $\la \neq 0$. The propagator of the $\vf$-field is $(q^{\si} + \la)^{-1}$ and the saddle-point equation reads
 \be \label{mf}  \frac{t}{\la} = \frac{2}{g} +  \frac{1}{(2\pi)^d}\int \frac{ d^dq}{q^\si (q^{\si}+ \la) }\ee
 For $d>2\si$ the integral converges when $\la$ vanishes and one obtains the mean field result
 \be \xi = \la^{-1/\si} \propto t^{-1/\si} \ee
 i.e. 
 \be \nu = \frac {1}{\si} \ee
 For $\si <d<2\si$ the integral in the r.h.s. of  (\ref{mf}) diverges near $T_c$ as $\la^{d/\si -2}$  i.e.
 \be \nu = \frac{1}{d-\si}\ee
 and from the scaling law $\nu = 1/(d-2+\eta)$ 
 one recovers the result
 \be \eta = 2-\si \ee
 in agreement with the fact that the critical two-point function is equal to $1/p^{\si}$. 
\end{itemize} \vskip 3mm
 \emph{The 1/N correction}\vskip 2mm
 The $\la$-propagator at $T_c$ is 
 \be \Delta_{\la}(p) = -\frac{2}{N} \left[ \frac{2}{g} + \int \frac{d^dq}{(2\pi)^d} \frac {1}{ (q^2)^{\si/2} ((p-q)^2)^{\si/2}}\right]^{-1} \ee
 and the integration, up to some UV cut-off $\Lambda$, gives for $p\ll \Lambda$ and $\si<d<2\si$ 
 \be \int \frac{d^dq}{(2\pi)^d} \frac {1}{ (q^2)^{\si/2} ((p-q)^2)^{\si/2}} = b(d,\si) p^{-(2\si-d)} \ee
 with 
 \be  b(d,\si)=\frac{1}{(4\pi)^{d/2}} \frac {\Gamma^2 (d/2-\si/2) \Gamma (\si -d/2)} {\Gamma^2 (\si/2) \Gamma (d -\si)} \ee
 
 The leading $1/N$ correction to the self-energy of the  $\vf$ critical propagator $1/p^{\si}$ is given by the exchange of one $\la$ field and one $\vf$ field between two vertices $\la \vf^{ 2}$. 
 
 Finally the inverse propagator of the $\vf$-field  is  is given by
 \be
 \label{G}
 \Gamma^{(2)} (p) = p^{\si} +  \frac{2}{N b(d,\si)}   \int \frac{d^d q}{(2\pi)^d} \left[  \frac{1}{ (q^2)^{d/2-\si}((p-q)^2)^{\si/2}}- \frac{1}{(q^2)^{(d-\si)/2}}\right]
 \ee
In a dimensional regularization scheme the zero-momentum subtraction term which ensures criticality vanishes automatically.
 In a cut-off theory the integral in the r.h.s.\ of (\ref{G}) would vanish as $p^2$ for small $p$, the standard non-renormalization of $\eta$. However in the dimensionally regularized, scale-invariant, theory the situation is not the same.
   The computation of the integral in the previous equation
follows standard techniques. One may use for instance Feynman's identity 
\be \frac {1}{A^{\alpha} B^{\beta} } = \frac {\Gamma(\alpha+\beta)}{\Gamma({\alpha}) \Gamma({\beta})} \int_0^{1} dx \frac{x^{\alpha-1} (1-x)^{\beta-1}} {[Ax+B(1-x)]^{\alpha+\beta}}\ee
to integrate over $q$. One finds
 \be \int \frac{d^d q}{(2\pi)^d}  \frac{1}{ (q^2)^{d/2-\si}((p-q)^2)^{\si/2}} = \frac{p^{\si}}{(4\pi)^{d/2}} \frac{\G({\si})\G(-\si/2)\ \G(d/2-\si/2)}{ \G(\si/2) \G(d/2-\si ) \G(d/2+ \si/2) }\ee
 Therefore, as explained in the previous section, one finds
 \be \G^{(2)} (p) = Zp^{\si} \ee
 with 
 \be Z = 1 + \frac{2}{N} \frac{\G (d-\si)\G({\si})\G(-\si/2)\G(\si/2)}{\G(d/2-\si/2 )\G(d/2+ \si/2) \G(\si-d/2)\G(d/2-\si) }+ O(1/N^2)\ee
 As announced previously, one sees that the cross-over from long-ranged to short-ranged interactions, namely the limit  in which $\si \to 2$ is singular since $\G(-\si/2)$ has a pole at $\si=2$  and $Z�^{-1}$ vanishes linearly with $(2-\si)$.  In the limit $\si \to 2$ one finds 
 \be Z^{-1} =  1- \frac{2}{N(2-\si)} \frac{(4-d) \G(d-2)}{\G^2(d/2-1)\G(2-d/2) \G(d/2 +1)}+ O(1/N^2) \ee
 
 In three dimensions for instance one finds for $\si$ close to 2:
 \be Z^{-1} =  1- \frac{8}{3N \pi^2} \frac{1}{(2-\si)} + O(1/N^2)\ee
  and since for the short-range problem \cite {SK}
  \be \eta_{SR} = \frac{8}{3N \pi^2}+ O(1/N^2)\,. \ee  
 $Z^{-1}$ vanishes when 
 \be 2-\si = \eta_{SR}\, , \ee
  
 This does holds in generic dimensions for large $N$  since \cite{SK}
 \be \eta_{SR} = \frac{2}{N} \frac{(4-d) \G(d-2)}{ \G^2(d/2-1) \G(d/2+1) \G(2-d/2)} +O(1/N^2) . \ee

In deriving these results we have made the implicit assumption that the two-loops contributions are divergent at most as
\be
\frac{1}{N^2(2-\si)}
\ee
and no double pole terms such as 
\be
\frac{1}{N^2(2-\si)^2}
\ee
 would appear in the computation of $\Gamma^{(2)} (p)$ at  next order. 
 
 In order to check this assumption we have to trace back the origin of the term $\frac{1}{N(2-\si)}$ in the one loop computation. 
 
 At one loop it is easy to check that in position space $\tilde\Gamma^{(2)} (x)$ is finite and the $(2-\si)^{-1}$ singularity arises from the Fourier transform. In other words the one-loop diagram, after subtraction at zero momentum, becomes superficially divergent when $(2-\si)\to 0$. This superficial divergence disappears after renormalization: the renormalized perturbative expansion does not contain divergences at the critical point. Therefore also all higher order contributions to $\tilde\Gamma^{(2)} (x)$ are finite and $\Gamma^{(2)} (p)$ is only superficially divergent and it can have only a $(2-\si)^{-1}$ singularity. Our implicit assumption is thus established.

\section{A numerical check: the case of percolation}

We want to provide a numerical example where we can check the validity of the proposed scaling relations.
We have chosen the two dimensional link percolation with long-range links for several reasons:
\begin{itemize}
\item Since we would like to observe logarithmic corrections, we need to study very large values of the system size $L$.
\item Usually we need a model in dimensions $d>1$ in order to have short-range models with non trivial exponents, but a low value of $d$ is convenient because the volume increases as $L^d$ and we want to study large $L$ values. The case $d=2$ is the best possible compromise. Moreover the value of $\eta$ for two dimensional percolation is known, i.e. $\eta=5/24\simeq 0.20833$ (indeed percolation correspond to the $n=1$ Potts model).
\item Percolation is the only problem that can be solved in a time (almost) linear in the system size by a simple algorithm. At variance with Monte Carlo methods, no thermalization is required and, for each choice of the random sample, the equivalent of the susceptibility can be computed exactly in linear time.
\end{itemize}

The problem is defined as follows.
We consider a square lattice of size $N = L \times L$ and we allow links to be placed between any pair of vertices with a probability that decays with the distance $r$ between the two vertices as $r^{-(d+\sigma)}$.
We use a notion of distance that decreases finite size effects due to boundary conditions \cite{APR}, while being equivalent to Euclidean distance in the $L \to \infty$ limit: the distance between sites $i$ and $j$ with coordinates $(i_x,i_y)$ and $(j_x,j_y)$ is given by
\[
r_{ij}^2 = \left[\sin\left((i_x - j_x) \frac{\pi}{L}\right) \frac{L}{\pi}\right]^2 +
\left[\sin\left((i_y - j_y) \frac{\pi}{L}\right) \frac{L}{\pi}\right]^2
\]
We study link percolation by varying the ratio $\alpha$ between the number of links and the number of sites (that we often call link density).

To the best of our knowledge this model has never been studied before; only some limits match known models: e.g. for $\sigma=-d$ the probability of having a link is distance-independent and we recover percolation in Erd\"os-R\'enyi random graphs; for $\sigma \to \infty$ we recover link percolation on the square lattice, where only nearest-neighbor links are allowed.

We have studied systems of size $L=2^\kappa$ with $\kappa \in [6,12]$ for several values of $\sigma \in [0,3]$; for $\sigma=\sigma^*=43/24=1.79167$ we have used sizes up to $\kappa=14$. The number of samples varies between $1.5\,10^5$ for $\kappa=6$ to roughly $7000$ for $\kappa=14$.

We have used the Hoshen-Kopelman algorithm which is able to keep an updated list of clusters, while adding links, in a time which is almost linear in the system size $N$. More details on our implementation of the algorithm can be found in Ref.~\cite{ScientificProgramming}.

In a percolation problem, correlations among pairs of variables are trivial, that is $C_{ij}=1$ if the sites $i$ and $j$ belong to the same cluster and $C_{ij}=0$ otherwise. For this reason a natural definition of the susceptibility is
\[
\chi \equiv \frac1N \sum_{i,j} C_{ij}\;,
\]
that should behave as follows at the critical point
\ba
\chi_c(L,\delta) \approx -\delta\, F^- L^\sigma \ & \text{for} & \ \delta < 0\\
\chi_c(L,\delta) \approx \frac{F^0 L^{\sigma^*}}{\log(L)} \ & \text{for} &\  \delta = 0\\
\chi_c(L,\delta) \approx \delta\, F^+ L^{\sigma^*} \ & \text{for} &\  \delta > 0
\ea
where $\delta=\sigma-\sigma^*$ and $\sigma^*=2-\eta_{SR}$.
The scaling law for the susceptibility we want to check is then
\be
\chi_c(L,\delta) \approx \frac{L^{\sigma}}{\log(L)} G[\delta\,\log(L)]\;,
\label{scalingChi}
\ee
where $G(z) = z F(z)$. The function $G(z)$ must be such that $G(z) = -F^-\,z$ for $z\to-\infty$, $G(0) = F^0$, and $G(z) = F^+\,z\,\exp(-z)$ for $z\to\infty$. A very simple form for this scaling function is
\be
G(z) = \frac{A\;z}{\exp(z)-1}\;,
\label{simpleF}
\ee
where $A=F^-=F^0=F^+$.

As long as there is no percolating cluster the susceptibility can be expressed in terms of clusters sizes as
\[
\chi = \frac1N \sum_{c=1}^C S_c^2\;,
\]
where $S_c$ is the size of the $c$-th cluster and $C$ is the number of clusters. That is $\chi$ is the average size of all clusters.
However, in presence of a percolating cluster of size $S_1$, the susceptibility would be essentially $S_1^2/N \sim O(N)$, and would be divergent in the thermodynamical limit.
In this case we have to proceed as for the computation of connected correlations and remove the dominant contribution given by the largest cluster, such that the susceptibility measures the average size of \emph{non percolating} clusters
\be
\chi = \frac{\sum_{c=2}^C S_c^2}{\sum_{c=2}^C S_c} = \frac{1}{N-S_1} \sum_{c=2}^C S_c^2\;.
\ee
In practice we use the latter expression for any value of $\alpha$. 

In the non percolating phase the previous definition of the susceptibility differs from the standard one by terms going to zero as one over the volume in the infinite volume limit. It is finite in the infinite volume limit away from the phase transition point and it is infinite only at that point.

\begin{figure}[t]
\begin{center}
\includegraphics[width=0.7\columnwidth]{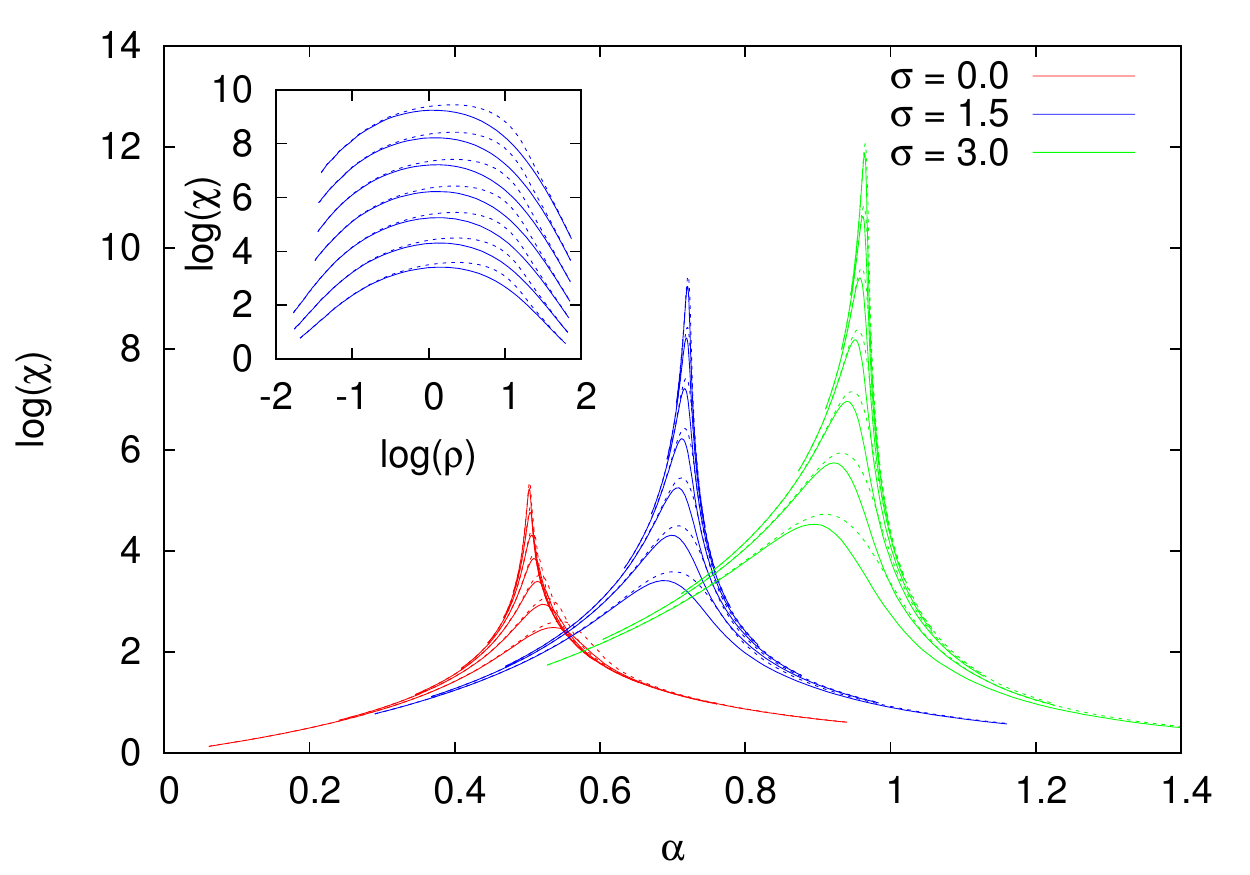}
\end{center}
\caption{Logarithm of the susceptibility as a function of link density for three different values of $\sigma$ for different system size. Full lines are $\overline{\log(\chi)}$ and dashed lines are $\log(\overline{\chi})$. In the inset, data for $\sigma=1.5$ are plotted versus the scaling variable $\rho = \overline{\log(S_1/S_2)}$. Upper curves are for larger sizes.}
\label{fig1}
\end{figure}

\begin{figure}[t]
\begin{center}
\includegraphics[width=0.58\columnwidth]{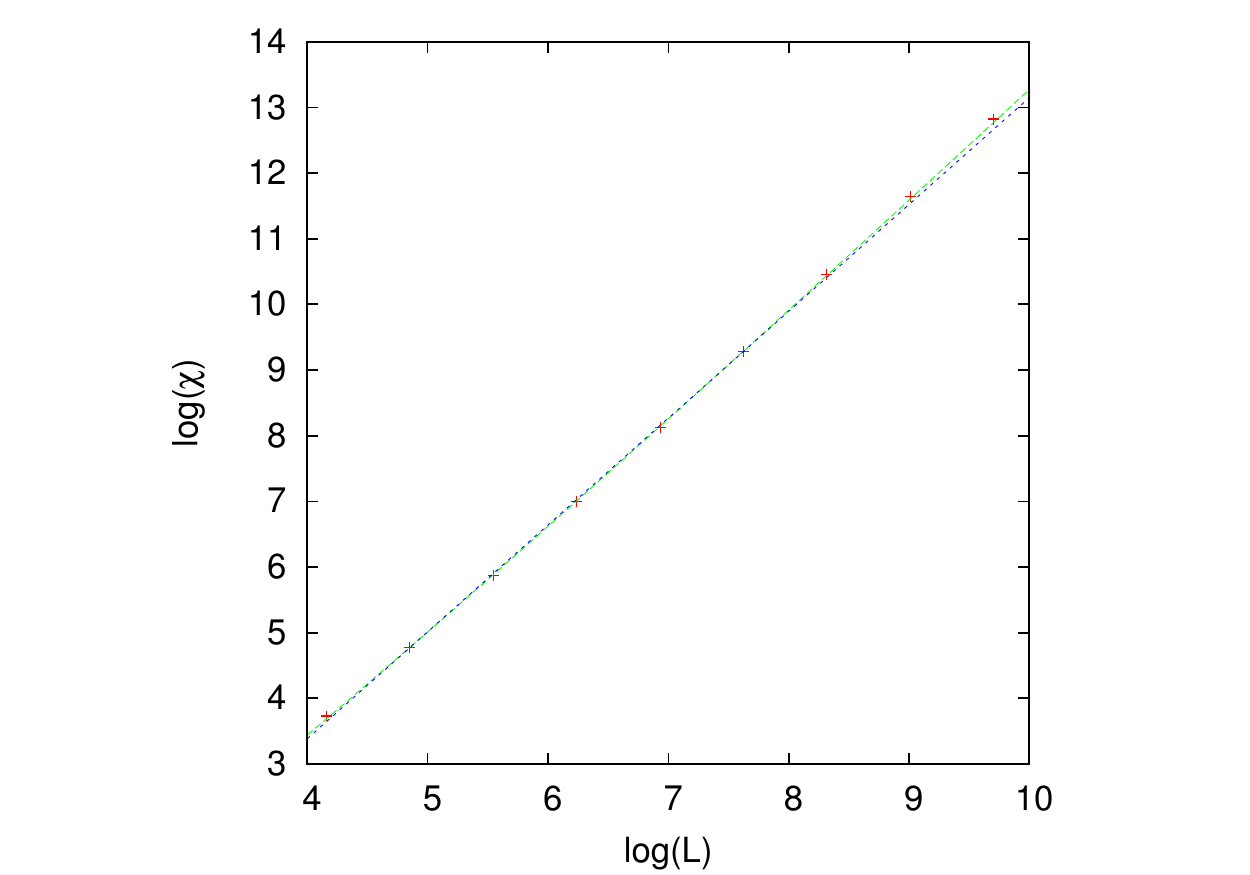}
\hspace{-3.3cm}
\includegraphics[width=0.58\columnwidth]{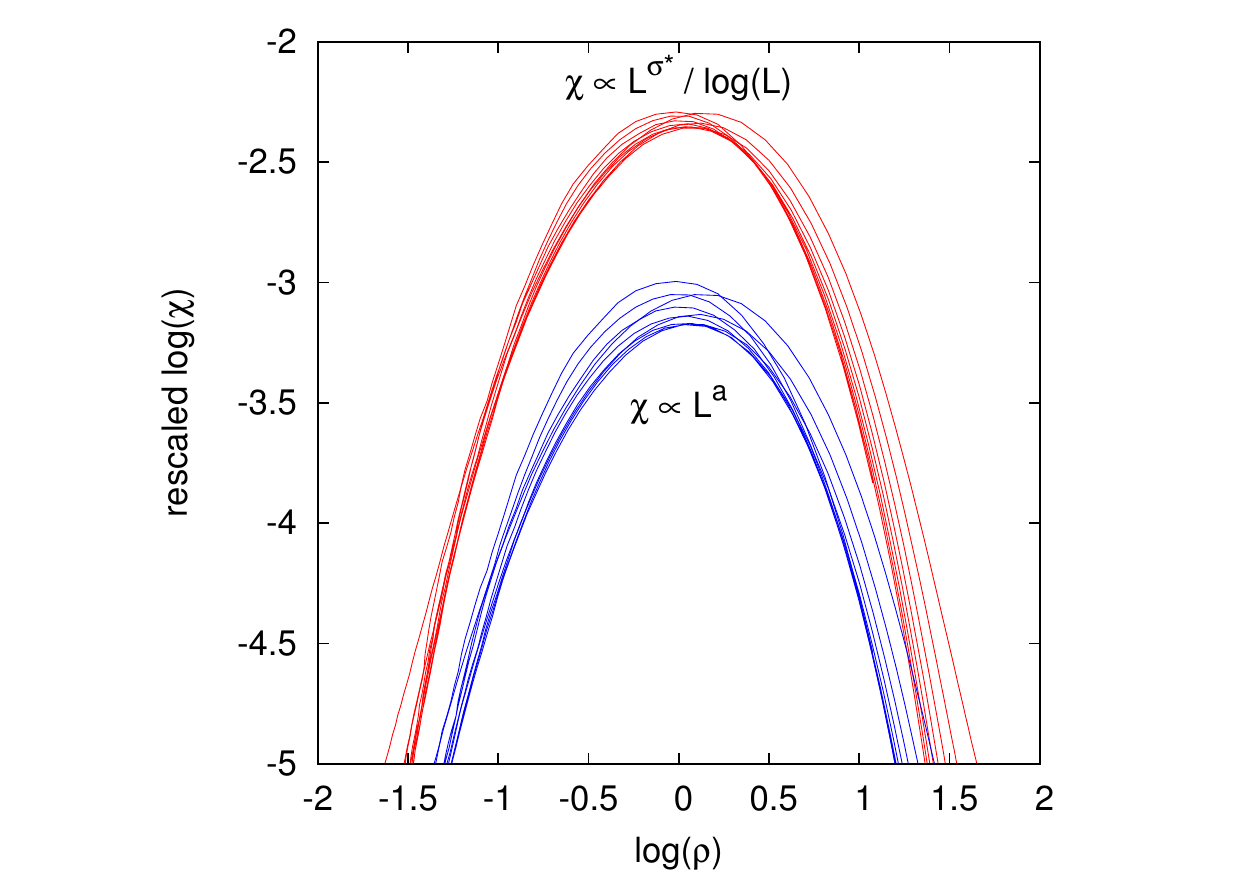}
\end{center}
\caption{Data at $\sigma=\sigma^*=2-\eta_{SR}$. Left: maximum of $\chi$ versus system size $L$ in a log-log scale; the green curve proportional to $L^{\sigma^*}/\log(L)$ fits data better than the best power law $L^a$(with $a=1.63$) shown with a blue dotted line. Right: scaling of the susceptibility in the whole critical region according to simple power law (lower blue curves) and with logarithmic corrections (upper red curves).}
\label{fig2}
\end{figure}

\begin{figure}
\begin{center}
\includegraphics[width=0.6\columnwidth]{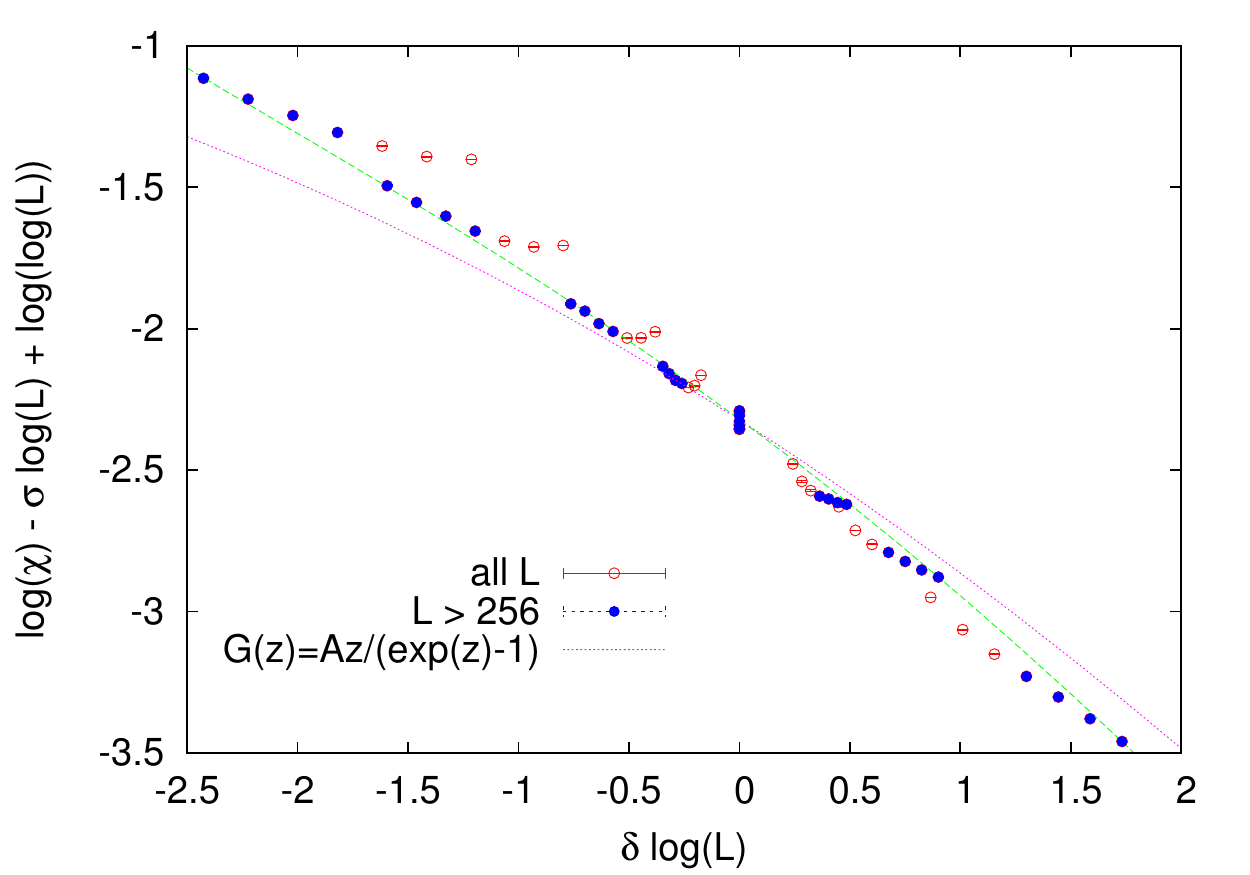}
\end{center}
\caption{A check of the scaling in Eq.(\ref{scalingChi}).}
\label{fig3}
\end{figure}

We are interested in studying how the susceptibility at the critical point grows with the system size.
Given that in disordered systems at the critical point, the observables typically show very large sample to sample fluctuations, we prefer to average the logarithm of the susceptibility in order to pick out the typical behavior.
In the main panel of Fig.~\ref{fig1} we show the logarithm of the susceptibility as a function of the link density $\alpha$ for three values of $\sigma$. Full lines are $\overline{\log(\chi)}$ and dashed lines are $\log(\overline{\chi})$, where the overbar is the average over the samples.
The behavior of the two averaged susceptibilities is very similar, the biggest difference being close to the critical point where sample to sample fluctuations are large.

In the inset of Fig.~\ref{fig1} we plot the same data of the main panel for $\sigma=1.5$ as a function of the scaling variable $\rho \equiv \overline{\log(S_1/S_2)}$, where $S_1$ and $S_2$ are respectively the largest and the second largest clusters. We notice that $\overline{\log(\chi)}$ is much more symmetric around the maximum with respect to $\log(\overline{\chi})$ and the location of the maximum of $\overline{\log(\chi)}$ is almost independent on the system size. For these reasons we compute the maximum of the logarithm of the susceptibility, which is a proxy for the critical susceptibility, by interpolating with a quartic polynomial the data of $\overline{\log(\chi)}$ versus $\log(\rho)$.

In Fig.~\ref{fig2} we provide a first evidence for the presence of logarithmic corrections at $\sigma=\sigma^*$. In the left panel we show the maximum of the susceptibility as a function of the system size: in a log-log scale it highlights the presence of a small upward curvature; indeed a simple power law fit $A+B \log(L)$ with 2 parameters does not interpolate well the data (blue dashed line), while a fit $A+\sigma^* \log(L)-\log(\log(L))$ with a single parameter does. On the right panel we show all the susceptibility data, not only the maximum, scaled according to a simple power law ($\chi \propto L^a$) or with logarithmic corrections ($\chi \propto L^{\sigma^*}/\log(L)$): the latter clearly provides a much better data collapse in the entire critical region  \emph{without any fitting parameter} (the outlier curves correspond to the smallest sizes, thus probably sensitive to scaling corrections).

In Fig.~\ref{fig3} we scale the data according to Eq.(\ref{scalingChi}) by plotting $\overline{\log(\chi)}-\sigma\log(L)+\log(\log(L))$ versus $\delta\,\log(L)$. We notice a rather good scaling, although some corrections to scaling are still visible: we draw in blue points corresponding to larger size ($L\ge 2^9$), which indeed lie closely to a common green curve. The purple curve is the very simple scaling function in Eq.~(\ref{simpleF}), with $A \simeq 0.1$.

\begin{figure}[t]
\begin{center}
\includegraphics[width=0.6\columnwidth]{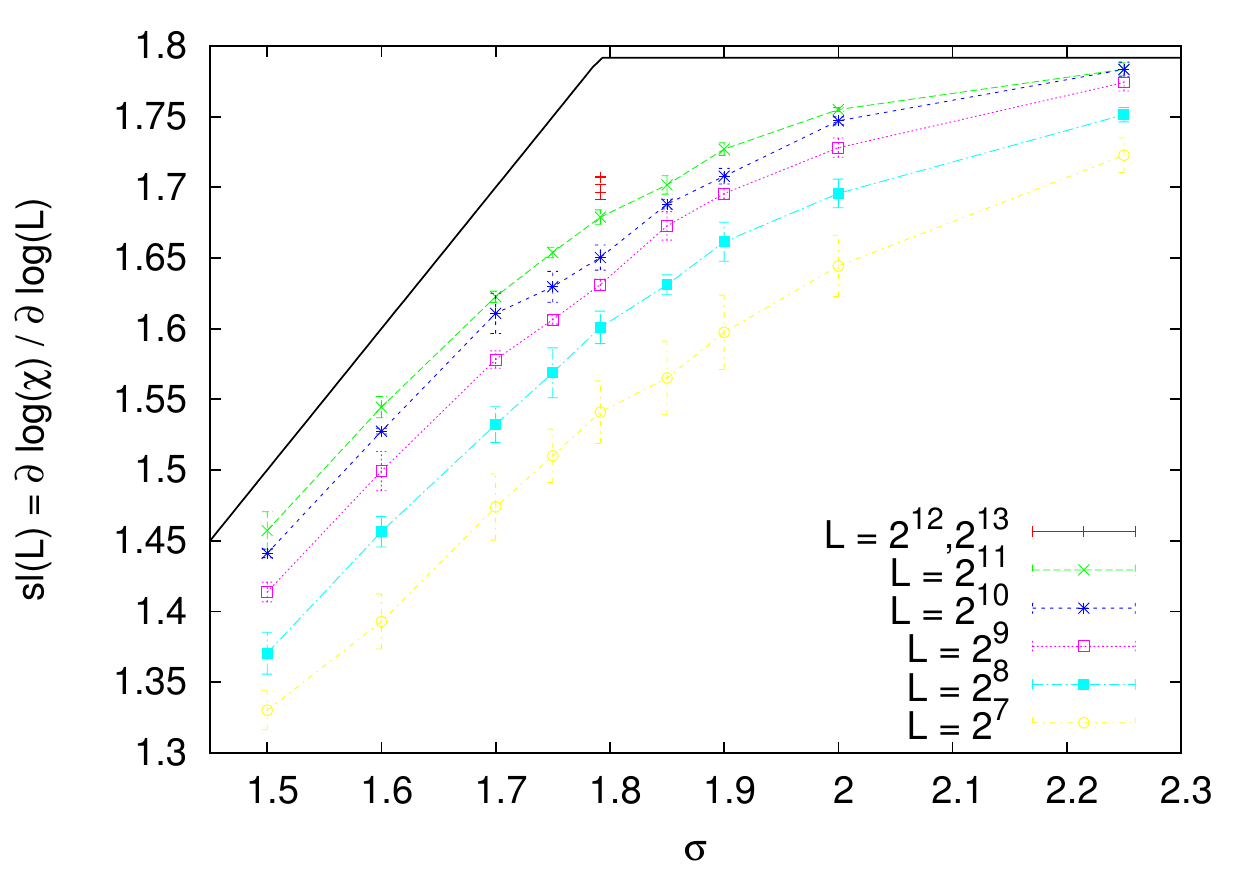}
\end{center}
\caption{Local slopes $sl(L) \equiv \partial\log(\chi)/\partial\log(L)$ converge to the asymptotic behavior (shown with black lines) with logarithmic corrections close to the crossover point $\sigma=\sigma^*=2-\eta_{SR}$.}
\label{fig4}
\end{figure}

\begin{figure}
\begin{center}
\includegraphics[width=0.6\columnwidth]{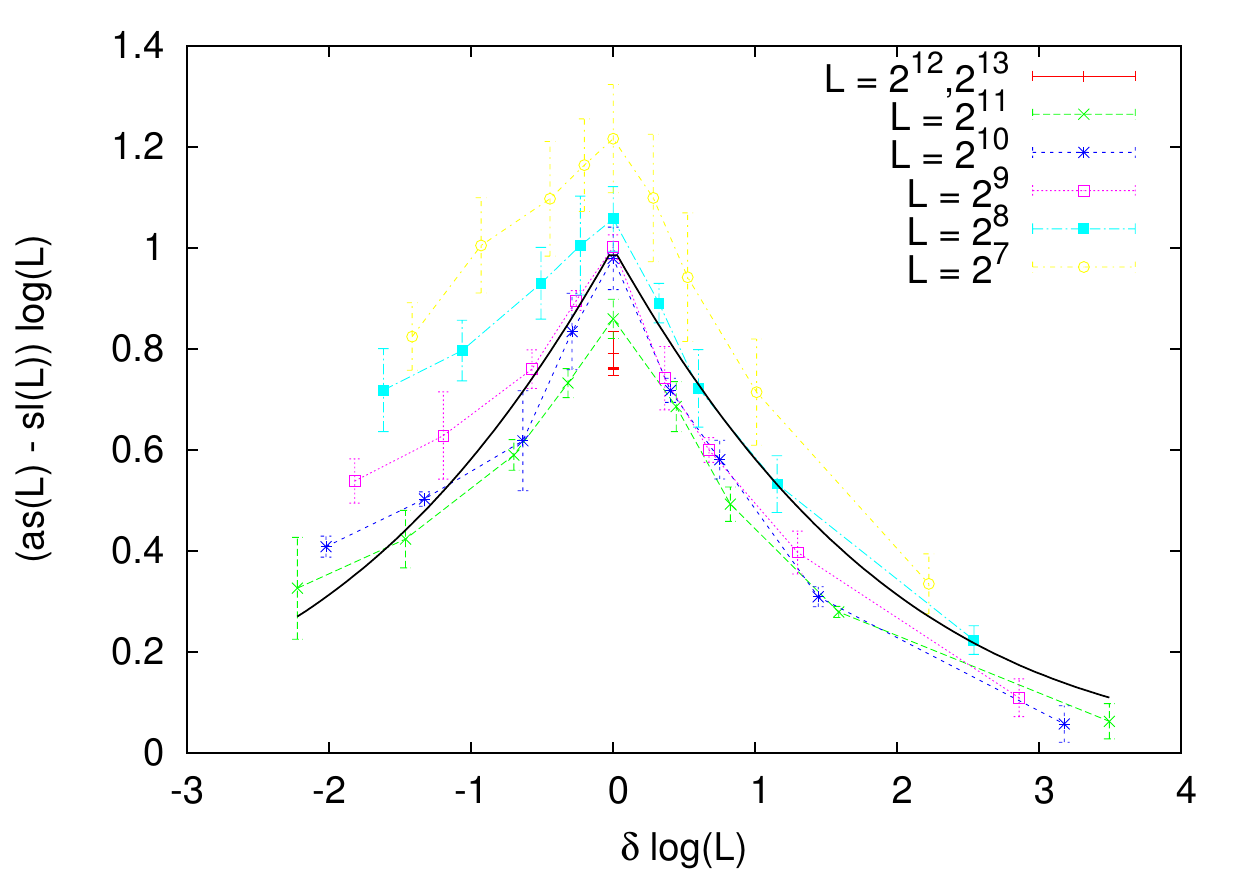}
\end{center}
\caption{Scaling of the local slopes $sl(L) \equiv \partial\log(\chi)/\partial\log(L)$ is consistent with the asymptotic exponent predicted by Sak, $as(\sigma) = \min(\sigma,\sigma^*)$, with logarithmic corrections at $\sigma=\sigma^*$, although smallest sizes still show correction to scaling. The black curve is the scaling function obtained with the simplest choice $F(z)=(\exp(z)-1)^{-1}$ and has no fitting parameters at all.}
\label{fig5}
\end{figure}

In previous works on the crossover region between long-range and short-range behaviors \cite{Picco} the exponent $\eta$ has been found to vary in a smooth way around $\sigma^*$, without any cusp. We believe this results may be due to the slow convergence to the asymptotic behavior. Indeed we have measured the local slope $sl(L) \equiv \partial\log(\chi)/\partial\log(L)$ by interpolating data at $L/2$, $L$ and $2L$. The results are shown in Fig.~\ref{fig4} and have the smooth behavior found by Picco \cite{Picco}. The black line is the asymptotic behavior, $as(\sigma) = \min(\sigma,\sigma^*)$.

Also the local slopes $sl(L,\sigma)$ may be rescaled according to the following scaling law
\be
sl(L,\sigma) = as(\sigma) - \frac{H(\delta\,\log(L))}{\log(L)}\;,
\ee
where the scaling function is given by
\[
H(z) = 1 - \frac{G'(z)\,z}{G(z)} - z\,\theta(z)\;,
\]
with $\theta(z)$ being the Heaviside step function.
In Fig.~\ref{fig5} we show the result of such a scaling, that looks acceptable although the smallest sizes have corrections to scaling. The black curve is the scaling function
\[
H(z) = \frac{z}{1-\exp(-z)} - z\,\theta(z)\;,
\]
obtained by the simplest choice $F(z)=(\exp(z)-1)^{-1}$ and describes the data reasonably well given that it has no fitting parameters at all.

\section{Conclusions}

We have found that   there is a non-trivial behaviour at the crossover point between short-range and long-range models. In particular at the crossover point we have logarithmic corrections to the standard power law behaviour.

The situation is similar to the behavior of the effective coupling constant for short-range model in $4-\epsilon$ dimensions where at the leading order in the region of small $k$ and $\epsilon$, we have.
\begin{equation}
g_R(k,\epsilon)\equiv k^\epsilon \Gamma_4(k,\epsilon)= \epsilon F(\epsilon \log(k))\, .
\end{equation}

We have looked for these logarithmic corrections in numerical simulations of two dimensional long-range percolation. We have  found that the critical exponents agree with the asymptotic theoretical predictions with  {\sl logarithmic} corrections. The logarithms are compatible with the scaling law  that we have proposed, although the range of the logarithm only changes, roughly speaking,  by  a factor 2 (i.e. from 4.5 to 9). Increasing the logarithm by another factor 2 is out of the range of present technology. More precise theoretical predictions, e.g. on the form of the crossover function, would be crucial to check the scenario that we propose.

{\it Acknowledgments -}  This research has received financial support from the European Research Council (ERC) through grant agreement No. 247328 and from the Italian Research Minister through the FIRB Project No. RBFR086NN1.

\end{document}